\documentclass{ws-procs961x669}            % default, citations in superscript
\usepackage{graphicx,color,ulem}
\usepackage{amsmath}
\def\gsim{ \lower .75ex \hbox{$\sim$} \llap{\raise .27ex \hbox{$>$}} }
\def\lsim{ \lower .75ex\hbox{$\sim$} \llap{\raise .27ex \hbox{$<$}} }

\newcommand{\ssr}{Space Sci. Rev.}
\newcommand{\nat}{Nature}
\newcommand{\aap}{A\&Ap}
\newcommand{\mnras}{MNRAS}
\newcommand{\apj}{Ap. J.} 
\newcommand{\apjl}{Ap. J. Lett.}
\newcommand{\jcap}{JCAP}
\newcommand{\prd}{Phys. Rev. D}
\newcommand{\araa}{Ann. Rev. Astron. \& Astroph.}

\begin{document}
\title{Who Ordered That? \\
 On The Origin of LIGO's Merging Binary Black Holes.}

\author{Tsvi Piran}

\address{Racah Institute of Physics, The Hebrew University of
		Jerusalem, Jerusalem 91904, Israel\\
E-mail: tsvi.piran@mail.huji.ac.il\\
https://scholars.huji.ac.il/tsvipiran/home}

\author{Kenta Hotekezaka}

\address{Department of Astrophysical Sciences,  Princeton University,
 Princeton, NJ 08544,  USA\\
E-mail: kentah@astro.princeton.edu}

\begin{abstract}
LIGO's detection of gravitational waves from binary black hole mergers was an unexpected surprise that immediately raised the question - 
 what is the origin of these black hole binaries? 
The ``simplest"  scenario is evolution of field massive stellar binaries. However, other possibilities involving capture have been proposed. 
We explore here one of the more interesting clues on this puzzle: the relatively modest spins of the resulting black holes that imply that the progenitor black holes were not spinning rapidly.  More specifically we consider the implication of observed distribution of, $\chi_{\rm eff}$,  the mass weighted projected  (along the orbital axis) spins on the field evolution scenario. 
In all cases $\chi_{\rm eff}$ is small and in two of the cases the best fit value is negative. 
Only in one event the spin is positive at 90\% credible.    These observations are puzzling within the field binary scenario in which positive higher spins ($\chi_{\rm eff} \ge 0.5$) are expected. At first sight one may expect that this rules out the field evolutionary scenario. 
Indeed we show that  
with typical parameters  a significant fraction ($\ge 25\%$) of the mergers should have high effective spin values. 
However,  uncertainties in the  outcome of the common envelope phase (the typical separation and whether the stars are rotating or not) and  in the late stages of massive star evolution  (the strength of the winds) make it impossible to rule out, at present,  these  scenarios. While observations of mergers with high effective spin  will support this scenario, future observations of negative spin mergers would rule it out. 
\end{abstract}


\bodymatter

\section{Prolog}
{\it Jacob Bekenstein is renowned for his deep insight and devotion to basic physics problems such as black hole's entropy, entropy bounds and TEVES. Less is known about his interest in astrophysics and his significant contributions to this field.  Naturally, Jacob has worked mostly, but not only, on the astrophysics of  black holes  and gravitational collapse. Among this work  Bekenstein 1973\cite{Bekenste73}, Bekenstein and Bowers 1974\cite{Bekenste74} and Bekenstein 1976\cite{bekenste76}  that deal with kicks give to black holes during gravitational collapse and with black holes in binary systems gained renewed significant forty years later with LIGO's discovery of merging black holes binaries, that we discuss here. }

\section{Introduction}\label{sec:Intro}
Gravitational-wave astronomy begun  in Sept 14th 2015 with LIGO's discovery \citep{abbott2016PhRvL} of  GW150914, a binary black hole (BBH) merger\footnote{ Indirect evidence for gravitational radiation was discovered in  the binary Pulsar \citep{hulse1975ApJ}, but LIGO's  was the first direct detection. }. 
Somewhat surprisingly this merger involved two massive black holes (36$m_\odot$ and 29$m_\odot $). 
An additional BBH merger, GW151226, as well as a merger candidate, LVT151012 were discovered in LIGO's O1 run. Three other events, GW170104 \citep{Abbott2017PRL}, GW170608 \cite{Abbott2017c} and GW170814 \citep{Abbott2017b}, the latter one jointly with Virgo,  were discovered in the O2 run.  Most of these BBHs  involved rather massive progenitor BHs that are larger than stellar mass black holes detected so far in X-ray binaries. The lightest black hole mass (observed in GW170608) is 7$m_\odot$. 

%With an estimated rate of $\sim 100$ events per year per cubic Gpc, BBH mergers are rare, but not extremely rare events. While at the level of 0.1\% of the rate of Supernovae this rate is comparable, for example, to the rate of Long Gamma-Ray Bursts \citep{wanderman2010MNRAS}. 

Among the most remarkable features of all six  events are the relatively low values (ranging between -0.12 and +0.21) of the mass weighted projected  (along the orbital axis) spins $\chi_{\rm eff}$ of the progenitors BHs. 
 In two cases the best fit values are negative (but the errors don't exclude zero), in three cases it is practically zero and only in one case the best fit value is positive but  even this value is small. These values are best fitted by  a low-spin isotropic distribution \cite{farr2017Nature,vitale2017}), suggesting that the progenitor black holes were not rotating rapidly and that their orientation was random. 
 
 This result is in 
 some ``tension"   with
the expectations  from  field binary evolution scenarios in which we expect that the individual spins should be both large and aligned with the orbital angular momentum axis. Namely,  we expect that  in this scenario a significant fraction of the mergers would have a large ($\gsim 0.5$)  $\chi_{\rm eff}$\citep{zaldarriaga2017,HotokezakaPiran2017ApJ}. 
The essence of the argument  that suggests a  large $\chi_{\rm eff}$ is the following \citep{kushnir2016MNRAS,HotokezakaPiran2017ApJ}:
\begin{romanlist}[(iii)]
\item  To merge within a Hubble time, $t_{\rm H}$,   the initial semi-major axis of the BBH at the moment of the formation of the second BH, $a$,  should be small.  Otherwise the gravitational waves merger time, that is proportional to $a^4$, would be too large. 
\item With a relatively small separation   the stars feel a significant tidal force and  their
spin tends to be synchronized with the orbital motion.  This is particularly relevant for the spin of the second star that collapses.  
\item  If synchronized,  the stellar spin, $S$,  is large relative to the maximal angular momentum of a black hole with the same mass, $Gm^2/c$.
that is $\chi_{*} \equiv S c/Gm^2 > 1 $. 
\item If there is no kick during the collapse, the progenitor's spin, $\chi_{*}$,  will determine the black hole's spin, $\chi_{BH}$. 
\item Hence we expect that  $\chi_{BH}$  will be large  and it will be oriented along the orbital axis. Therefore   $\chi_{*}(t_{\rm H})/2 <\chi_{\rm eff} \lsim 1/2$  if only the secondary  has been synchronized and $ \chi_{*}(t_{\rm H})< \chi_{\rm eff} \lsim 1$ if both progenitors have been synchronized. $\chi_{*}(t_{\rm H})$ is the spin ratio of a star in a binary system that merges within a Hubble time\footnote{For simplicity we assume in most, but not all,  of the discussion that both black holes have  similar masses. %Note also that if angular momentum is conserved but mass is lost during the collapse of the progenitor star to a black hole $\chi_{BH}$ will be larger than $\chi_{*}$.
}.
\end{romanlist}

The most favorable ``standard"\footnote{{\bf Mixed star scenario.}}  evolutionary scenarios involves Wolfe-Rayet (WR) stars. These are  massive stars that lost their envelopes. 
We compare here the observed distribution with those expected in this scenario.  
We begin  in \S \ref{sec:obs} with a discussion of the gravitational wave observations. We turn in \S \ref{sec:galactic} to   observations of  Galactic X-ray binaries containing BHs. In \S \ref{sec:theory},
we  express the initial semi-major axis, $a$, in terms of the the merger time, $t_c$,
and we estimate  $\chi_*(t_{c})$ in terms of the progenitor's parameters  $t_c$.  We include in these calculations also the effect of angular momentum loss via winds during the late stages of the stellar evolution.  
In \S \ref{sec:collapse} we discuss possible changes in the spin during the collapse to the black hole. 
In \S \ref{sec:results},  we calculate, using these estimates,  the expected spin distribution in different scenarios and compare it to the gravitational-wave observations. We conclude in \S  \ref{sec:summary}   and summarize the results. We then address the question raised earlier  whether the observations disfavor  field binary evolution models \citep{belczynski2016Nature,belczynski2017,stevenson2017NatCo,postnov2017arXiv,OShaughnessy17}  and  support  capture models  \citep{ioka1998PRD,rodriguez2016ApJb,oleary2016ApJ,
antonini2016ApJ,bartos2017ApJ,sasaki2016PhRvL,bird2016PhRvL,blinnikov2016JCAP,kashlinsky2016ApJ,Rodriguez2016a}, in which the spins are expected to be randomly oriented. 

%The discussion follows to a large extent  \citep{HotokezakaPiran17} \citep[see also][for a related discussion]{kushnir2016MNRAS,zaldarriaga2017} . 
%We note that while the apparent ``tension" between the expectations arising from these arguments was already apparent in February 2017 when this talk was given, the ``tension" increased with the announcement of the observations of GW170104, announced in early June.  

\section{Binary BH Mergers Observations} 
\label{sec:obs}
\begin{figure}[h]
\vskip -1.5cm
  \begin{center}
    \includegraphics[bb = 0 0 410 302, width=0.8\linewidth]{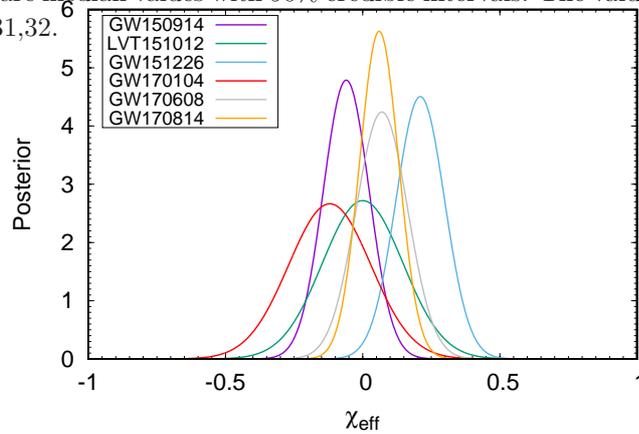}\\
\vskip -0.5cm
    \caption{The distribution of the observed spins. We have approximated each observed distribution as a Gaussian whose mean value and $90\%$ credible interval are the  values shown in 
     Refs. \citenum{abbott2016PhRvX,Abbott2017PRL,Abbott2017dd,Abbott2017b,Abbott2017c} (see also Ref. \citenum{farr2017Nature}). } 
    \label{fig:obsspins}
  \end{center}
\end{figure}

The relevant  observed properties of the BBH merger events are summarized in Table I. The most interesting ones for our purpose are the BHs' masses and their  $\chi_{\rm eff}$ values.
This latter quantity is  defined as: 
 \begin{equation} 
 \chi_{\rm eff} \equiv \frac{ m_1 \chi_1 + m_2 \chi_2 }{m_{\rm tot}}  \ ,
 \end{equation}  where 
 \begin{equation} 
 \chi_{1,2} \equiv \frac{c\vec S_{1,2} \cdot \hat L } { G m_{1,2}^2}\ , 
\end{equation}
and $m_{tot}=m_1+m_2$ and $\hat L$ is a unit vector in the direction of the system's orbital angular momentum $\vec L$.
The limits on $\chi_{\rm eff}$ are obtained from the observations of the gravitational-wave signals before and after the merger. 
In particular the lack of extended ringdown phases  puts  limits on the spins of the final BHs, $a_f$. 
 The fact that $a_f \sim 0.6-0.7$ is an independent evidence that the initial aligned spins of the BHs were not close to unity \citep{rezzolla2008ApJ,rezzolla2008PhRvD}.
Had the initial aligned spins been large, the final spin of the merged BHs would have been very close to unity and the GW signal would  have had a long ringdown phase.  
Indeed, the final spin is slightly larger ($0.74^{+0.06}_{-0.06}$) for GW151226, 
%($0.70^{+0.07}_{-0.05}$) for GW170814, and ($0.06^{+0.12}_{-0.12}$) for GW180914, the  cases 
for which the $\chi_{\rm eff}$ is largest. 
Fig. \ref{fig:obsspins}  describes the observed $\chi_{\rm eff}$ distribution in terms of  the corresponding six  Gaussians corresponding to the posterior distributions of the observed events.% and the resulting combined spin distribution for the whole sample.   

\begin{table*}[h]
\begin{center}
%\vskip 0.5cm
\label{tab:LIGO}
%\scalebox{1.}
{\begin{tabular}{lcccccc}
\hline \hline
Event & $m_1$           & $m_2$                 & $m_{\rm tot}$ & $\chi_{\rm eff}$ & $a_f$  \\%& Rate  \\
          & $[m_{\odot}]$ &   $[m_{\odot}]$    &$[m_{\odot}]$  &            &          \\  \hline  %&    [Gpc$^{-3}$ yr$^{-1}$]                
GW150914 & $36.2^{+5.2}_{-3.8}$ & $29.1^{+3.7}_{-4.4}$ & $65.3^{+4.1}_{-3.4}$ & $-0.06^{+0.14}_{-0.14}$ &$0.68^{+0.05}_{-0.06}$ \\%& $3.4^{+8.6}_{-2.8}$\\
GW151226 & $14.2^{+8.3}_{-3.7}$ & $7.5^{+2.3}_{-2.3}$ & $21.8^{+5.9}_{-1.7}$   & $0.21^{+0.20}_{-0.10}$ &$0.74^{+0.06}_{-0.06}$ \\%& $37^{+92}_{-31}$\\
LVT151012 & $23^{+18}_{-6}$ & $13^{+4}_{-5}$ & $37^{+13}_{-4}$                         & $0.0^{+0.3}_{-0.2}$ &$0.66^{+0.09}_{-0.10}$ \\%& $9.4^{+30.4}_{-8.7}$\\
GW170104 & $31.2^{+8.4}_{-6.0}$ & $19.4^{+5.3}_{-5.9}$ & $50.7^{+5.9}_{-5.0}$  & $-0.12^{+0.21}_{-0.30}$ & $0.64^{+0.09}_{-0.20}$ \\%& --\\
GW170608 & $12^{+7}_{-2}$ & $7^{+2}_{-2}$ & $19^{+5}_{-1}$  & $0.07^{+0.23}_{-0.09}$ & $0.69^{+0.04}_{-0.05}$ \\%& --\\
{GW170814} & $30.5^{+5.7}_{-3.0}$ & $25.3^{+2.8}_{-4.2}$ & $55.9^{+3.4}_{-2.7}$  & $0.06^{+0.12}_{-0.12}$ & $0.70^{+0.07}_{-0.05}$ \\%& --\\
\hline \hline 
\end{tabular}}
\end{center}
%\caption{Parameters of the BBH mergers detected during LIGO's O1 and O2 runs. The parameters are median values with 90\% credible intervals. 
% The values are taken from Ref. \citenum{abbott2016PhRvX,Abbott2017PRL}.}
{Table. 1: Parameters of the BBH mergers detected during LIGO's O1 and O2 Run. The parameters are median values with 90\% credible intervals. The values are taken from Refs. \citenum{abbott2016PhRvX,Abbott2017PRL,Abbott2017dd,Abbott2017b,Abbott2017c}.}
\end{table*}

\section{Galactic  BHs in X-ray binaries}
\label{sec:galactic}
 Observations of X-ray binaries involving BHs, albeit smaller mass ones, can also shed some light on the problem at hand.  In particular observations of two such systems that include massive ($>10 m_\odot$) BHs, Cyg X-1 and GRS 1915+105,   provide a good evidence that these massive BHs formed in situ,  in a direct implosion and without a kick \citep{mirabel2016arxiv}.  For example, Cyg X-1 moves at $9 \pm 2$ km/s relative to the stellar association Cygnus OB3, indicating that it could have lost at most $1 \pm 0.3 m_\odot$ at BH formation. 
Furthermore, the minuscule eccentricity of Cyg X-1, $0.018 \pm 0.0003$,   \citep{orosz2011ApJ}  suggests that the orbit has been circularized during the binary evolution and the collapse didn't give the system a significant kick that disturbed the circular orbit. {On the other hand,  low mass BH binaries, GRO J1655-40 ($m_{\rm BH}=5.3\pm0.7m_{\odot}$) and V404 Cyg ($m_{\rm BH}=9.0\pm0.6m_{\odot}$), have  larger peculiar velocities of $112\pm13$\,km/s and $39.9\pm5.5$\,km/s respectively \citep{mirabel2016arxiv}. These last two observations suggest  a  natal kick prescription of $v_{k,{\rm BH}}\approx m_{\rm ns}v_{k,{\rm ns}}/m_{\rm BH}$  for low mass BHs, where $m_{\rm ns}\approx 1.4m_{\odot}$ and $v_{k,{\rm ns}}\approx 300$\,km/s \citep{janka2013MNRAS}.}
 {In addition, Ref. \citenum{mandel2016MNRASb} shows that large ($>80$ km/s) natal kicks  are not required to  explain the observed positions of low-mass X-ray binaries in the Galaxy.  }

{
A natal kick can change the orbital parameters of the binary. 
For instance, 
the misalignment angle between the final orbital angular momentum vector and
the spin vectors, that are assumed to be aligned to the initial orbital angular momentum vector,   is given by (e.g. Ref. \citenum{tauris2017ApJ}):
\begin{eqnarray}
\delta = \tan^{-1}
\left( \frac{v_{k,{\rm BH}}\sin \theta \sin \phi}{\sqrt{(v_{\rm rel} +v_{k,{\rm BH}}\cos \theta)^2 + (v_{k,{\rm BH}} \sin \theta \cos \phi)^2}}\right),
\end{eqnarray}
where $\theta$ is the angle between the kick vector and the initial orbital plane,
$\phi$ is measured in the plane perpendicular to the orbital velocity vector of the collapsing progenitor star, and
$v_{\rm rel}$ is the relative velocity between the two stars:
\begin{eqnarray}
v_{\rm rel} = 680\,{\rm km/s}\,
\left(\frac{1+q}{2 q^{2/3}}\right)^{3/8} 
\left(\frac{t_c}{1\,{\rm Gyr}}\right)^{-1/8}
\left(\frac{m_2}{30m_{\odot}}\right)^{1/8},
\end{eqnarray}
where $q\equiv m_2/m_1$.
This is larger than the magnitude of the natal kicks observed in  Galactic X-ray binaries, 
thereby it is unlikely that BH natal kicks induce a large misalignment between the
orbit and the spin vectors.   
In the case of $v_{k,{\rm BH}}\ll v_{\rm rel}$, the misalignment angle, 
inferred from the estimates of the natal kicks of the Galactic BHs,
is $\delta \lsim v_{k,{\rm BH}}/v_{\rm rel}\lsim 0.15$.
Therefore we conclude that BH natal kicks could affect the misalignment between the orbit and spins
only if the natal kicks of BBH-merger progenitors are significantly larger than those observed in 
Galactic BH binaries. 
}

Estimates of the spins of the  BHs  in Cyg X-1\citep{McClintock2014SSRv} and GRS 1915+105\cite{Fragos2015} suggest that in these two systems  $a/m>0.95$. Three other BHs, LMC X-1, M33 X-7, and 4U 1543-47, whose masses are larger than $9 m_\odot$, have $\chi>0.8$.  Only one BH with a mass $> 9 m_\odot$, XTE J1550-564 has a significantly lower value ($\chi = 0.34^{+.20}_{-.28}$).  It is important to note that these large spins must be obtained at birth as accretion cannot spin up a massive BH to such a high spin value \citep{Valsecchi10,Wong14}. 

To summarize, the massive Galctic BHs observed in X-ray binaries  have large spins and were formed with no significant kicks. As these BHs are  clearly a part of  evolutionary systems  this suggests that other BHs that form is such systems will also have large spins.

\section{Merger Time, Orbital Separation and Synchronization. }
\label{sec:theory}

The merger time, due to gravitational radiation driven orbital decay, 
is: 
\begin{equation}
t_c  %= \frac{5}{256} \frac{a}{c}\frac{c^2a}{Gm_1}\frac{c^2a}{Gm_2}\frac{c^2a}{Gm_{\rm tot}} 
\approx 10~ {\rm Gyr}~ \left(\frac{2 q^2}{1+q}\right)\left(\frac{a}{44R_{\odot}}\right)^4
\left(\frac{m_2}{30M_{\odot}}\right)^{-3}\ ,
%\left(\frac{m_2}{30M_{\odot}}\right)^{-1}
%\left(\frac{m_{\rm tot}}{60M_{\odot}}\right)^{-1} \ .
\end{equation}
where $q=m_2/m_1$ is the mass ratio.
Note that  we assume circular orbits here and elsewhere. This simplifying assumption is based on the expectation that the orbit will be circularized during the binary evolution and that it won't be affected by the collapse of the secondary. It is supported by the observations of Galactic binaries containing massive BHs, discussed earlier (see \S \ref{sec:galactic}).

Tidal forces exerted by the primary, denoted by the subscript 1,  will tend to synchronize the secondary star, denoted by the subscript 2.  
If fully synchronized the final stellar dimensionless spin (normalized to the maximal spin of a BH with the same mass) would be: 
\begin{equation}
\label{Eq:spin}
\chi_2 \approx 0.5~
q^{1/4}\left(\frac{1+q}{2} \right)^{1/8} \left(\frac{\epsilon}{0.075}\right)   \left(\frac{R_2}{2R_{\odot}}\right)^{2} 
\left(\frac{m_2}{30M_{\odot}}\right)^{-13/8} \left(\frac{t_c}{1\,{\rm Gyr}}\right)^{-3/8}  \ ,
\end{equation}
where $\epsilon \equiv I_2  / m_2 R_2^2$ characterizes the star's moment of inertia. 
The progenitor's spin,  $\chi_{2}$, increases with the progenitors size and decreases when $t_c$ increases.  
%Thus, a compact progenitor star that formed at a high redshift produces  a low spin BH, while a large progenitor  formed recently collapses to a large spin BH \cite{kushnir2016MNRAS}. 

The synchronization process takes place over $t_{\rm syn}$:
\begin{equation}
t_{\rm syn} \approx  20~{\rm Myr}~\frac{(1+q)^{31/24}}{q^{33/8}} 
\left(\frac{\epsilon}{0.075}\right) \left(\frac{E_{2}}{10^{-5}}\right)^{-1}
\left(\frac{R}{2 R_{\odot}}\right)^{-7}  
%\nonumber \\ &&
\left(\frac{m_2}{30M_{\odot}}\right)^{47/8}
\left(\frac{t_c}{1 {\rm Gyr}}\right)^{17/8} \ ,
\label{eq:Zahn}
\end{equation}
%(131072 G^(51/8) m2^(47/8) (1 + q)^(17/8) TC^(17/8))/(25 5^(1/8) c^( 85/8) q^(17/8))
% {\bf I have changed from the original text $14 R_\odot$ to $7 R_\odot$ to correspond to a WR star. } 
 %\notekh{Note: the radius of a WR star with $30M_{\odot}$ is $\sim 2R_{\odot}$.}
where $E_2$, is a dimensionless quantity \cite{zahn1975A&A}
characterizing the inner structure of the star.  
$E_2$ is  $\sim 10^{-7}$--$10^{-4}$ for 
massive main sequence stars and Wolf-Rayet (WR) stars \citep{zahn1975A&A,kushnir2017MNRAS}. 
The characteristic values used in Eq. (\ref{eq:Zahn}) correspond to a WR star. 
For WR stars,  one can show \citep{kushnir2016MNRAS} that $t_{\rm synWR} $ is almost independent of $M_2$ and it can be expressed as:
\begin{eqnarray}
t_{\rm synWR} \approx 10~{\rm Myr}~q^{-1/8}\left(\frac{1+q}{2q}\right)^{31/24}\left(\frac{t_c}{1~{\rm Gyr}}\right)^{17/8}. 
\label{eq:synWR}
\end{eqnarray}

Because of their short stellar lifetime, WR stars are not necessarily synchronized in binary systems even with $t_c$ of a few hundreds Myr. 
In addition angular momentum can be lost due to winds during the last phases of the evolution of the star. 
Therefore the final stellar spin depends   on: 
\begin{romanlist}[(iii)]
\item
$\chi_i$, the spins of the stars at the beginning of the WR phase;
\item   the ratio of   $t_{\rm syn}$ and the lifetime of the WR star, $t_{\rm WR}$; 
\item  the  angular momentum 
loss timescale during the WR phase, $ t_{\rm wind}$ (see \citenum {kushnir2016MNRAS,HotokezakaPiran2017ApJ}).
\end{romanlist}

We characterize the angular momentum loss due to winds by defining  $t_{\rm wind}\equiv J_s/\dot{J}_s$, where $J_s$ is the spin angular momentum of the star and $\dot{J}_s$ the angular momentum loss rate. 
Stronger winds corresponding to shorter $t_{\rm wind}$ values.
To take account both of the stellar wind  and the circularization process during the  short stellar life time we solve the following equation to 
obtain the stellar spin parameter at the end of the WR phase \citep{kushnir2016MNRAS}:
\begin{eqnarray}
\dot{\chi}_* = \frac{\chi_{\rm syn}}{t_{\rm syn}}\left(1-\frac{\chi_*}{\chi_{\rm syn}} \right)^{8/3} - \frac{\chi_*}{t_{\rm wind}},
\end{eqnarray}
where $\chi_{\rm syn}$ is the stellar spin parameter in the synchronized state.

\section{Collapse  and  the BH Spin}
\label{sec:collapse}
One can expect that, unless there is too much angular momentum (that is $\chi_* \le 1$),   
the collapsing star implodes and the BH that forms  swallows all the collapsing stellar mass\footnote{The original stellar mass could be larger but this lost in an earlier phase due to winds \citep{mirabel2016arxiv}.}. If $\chi _*> 1$  a fraction of the matter will be ejected carrying  the excess angular momentum and  leading to a BH with $\chi \le 1$  \citep{stark1985PhRvL,oconnor2011ApJ,sekiguchi2011ApJ}. Thus we expect that 
\begin{equation}
\chi_{\rm BH} \approx  
\begin{cases}
 1  & \mbox{if } ~\chi_* \ge 1 ,  \\
\chi_*  &  \mbox{if }  ~\chi_* < 1 . 
\end{cases}  
\end{equation}

One may wonder if there are caveats to this conclusion. First, is it possible that matter
is ejected during the collapse to a BH even if 
 $\chi_* < 1$? This will, of course,
change the relation between the progenitor's spin and the BH's spin. Second is mass ejected isotropically? 
If not the BH will receive a kick and the BBH will be put into an elliptical orbit (that will merge faster)\cite{Bekenste73}. The kick may
also change the resulting BH spin. Since the initial spin is in the direction of the
orbital angular momentum the kick may reduce the spin component along this direction. Clearly these issues can only be addressed by a
detailed numerical study of the collapse. However, as discussed in \S \ref{sec:obs} observations of  Cyg X-1 and GRS 1915+105. Galactic binaries containing massive  ($>10 m_\odot$) BHs, provide a good evidence that massive BHs form in situ  in a direct implosion and without a kick  \citep{mirabel2016arxiv}.  
Estimates of the spins of accreting massive BHs give an independent support to this conclusion.

\begin{figure*}
  \begin{center}
        \includegraphics[width=6cm]{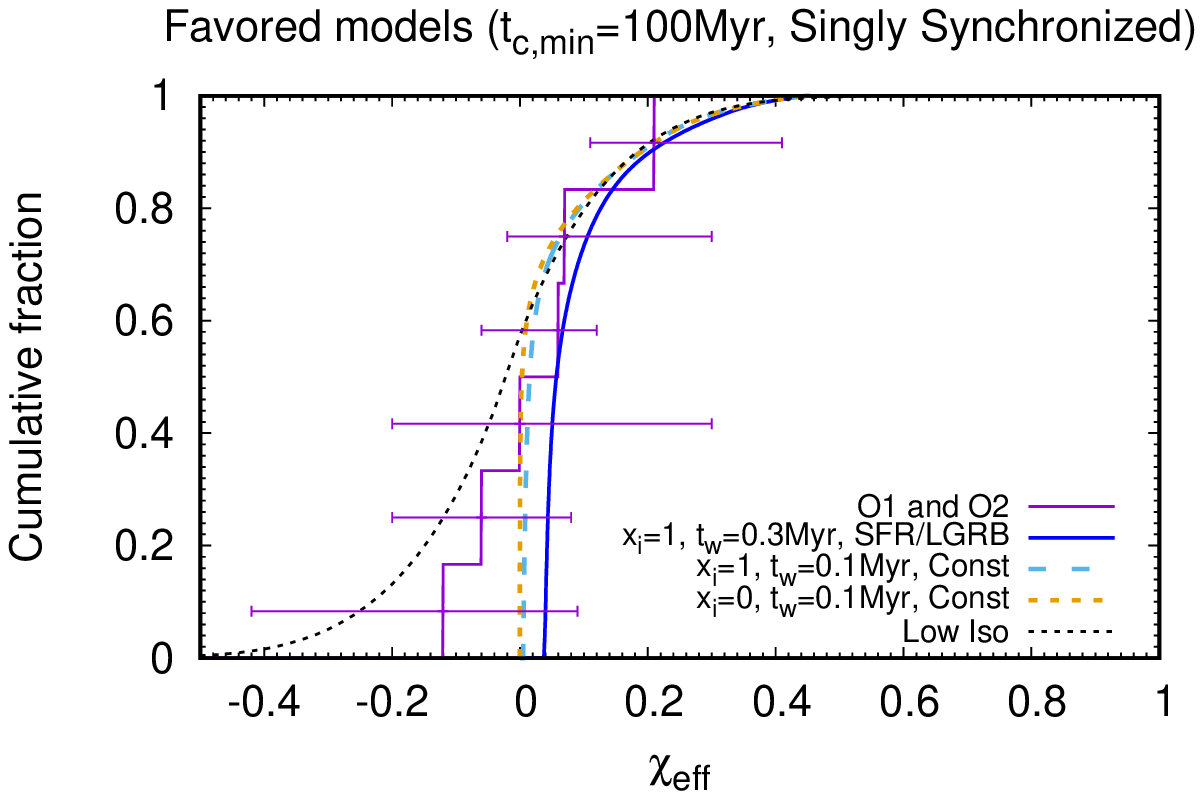}
         \includegraphics[width=6cm]{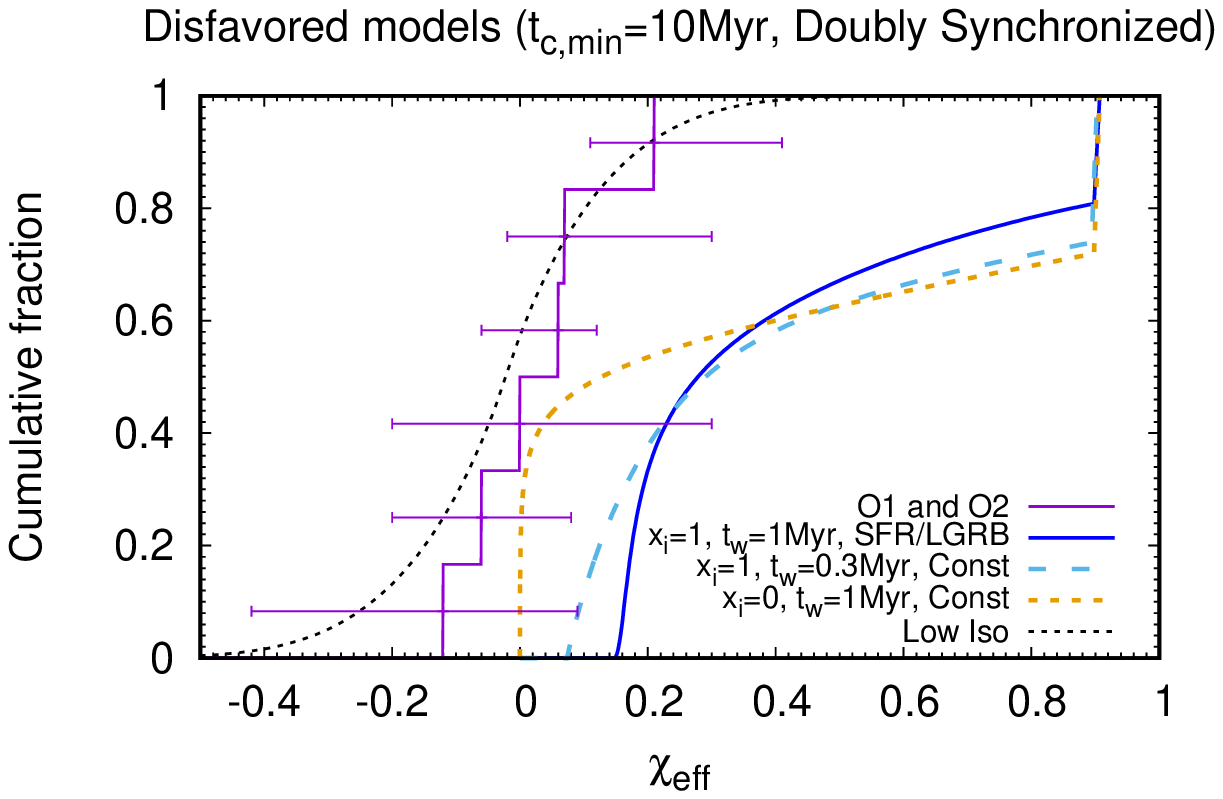}
           \caption{The cumulative $\chi_{\rm eff}$ distribution  for the O1 and O2 observing runs   for the favored ({\it left}) and disfavored ({\it right}) models for WR binary scenario.
   We set the mass ratio, $q=1$, and  $m_{\rm tot} = 60M_{\odot}$ for all models.  Also shown as a black dashed curve is the low-spin isotropic model of Ref. \citenum{farr2017Nature}. %   Note also that the theoretical curves are  calculated for BBHs with a total mass of $60M_{\odot}$. 
}
    \label{fig:KSbest}
  \end{center}  
\end{figure*}

\begin{figure*}
  \begin{center}
        \includegraphics[width=6cm]{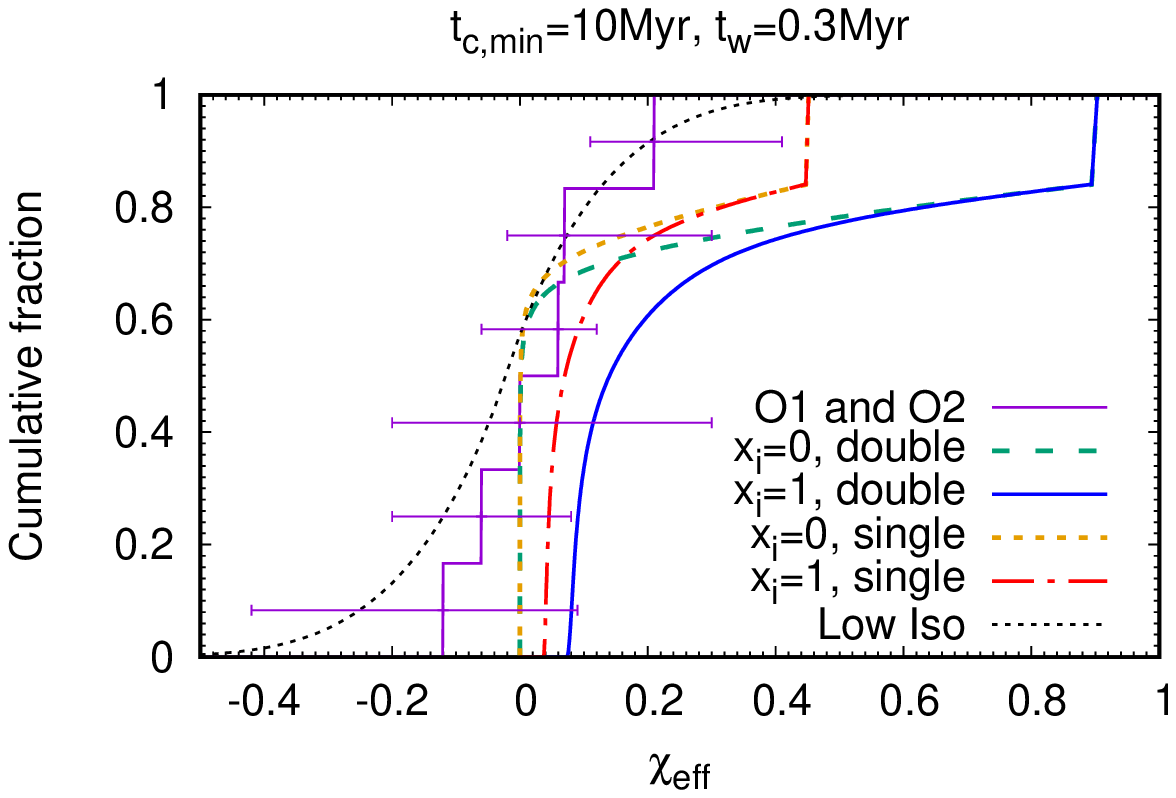}
         \includegraphics[width=6cm]{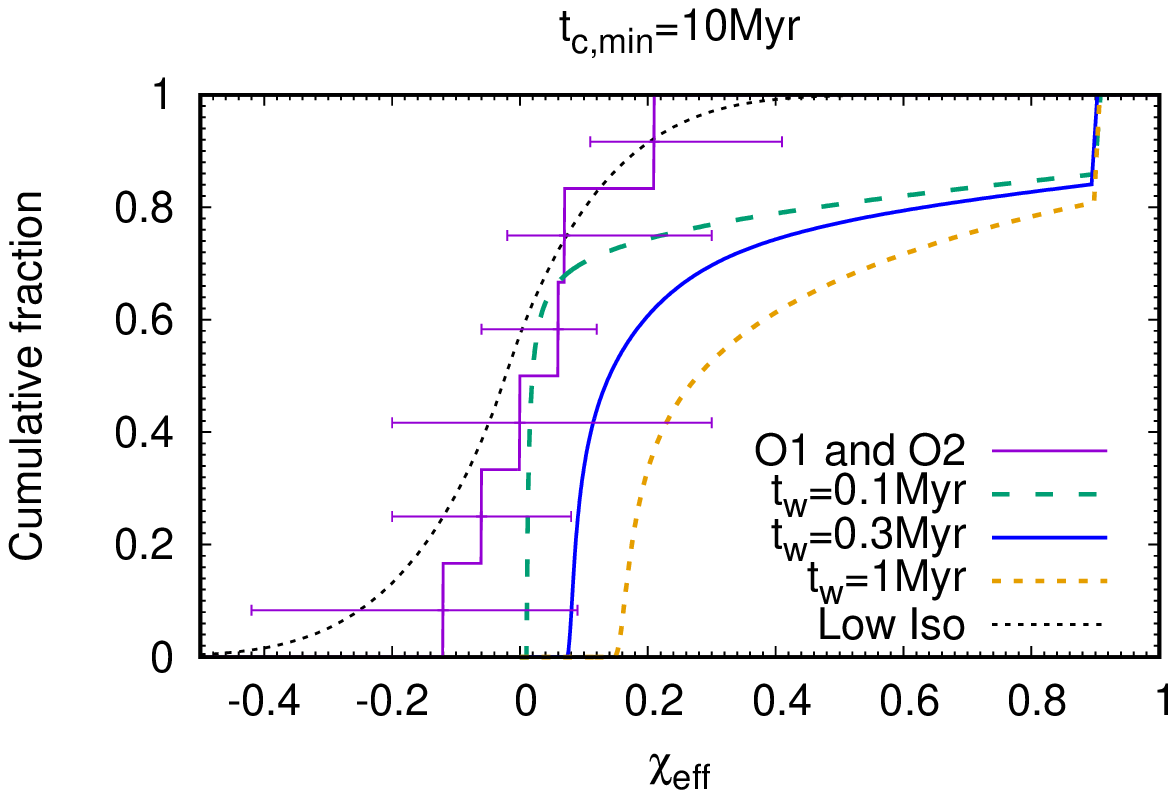}\\
                 \includegraphics[width=6cm]{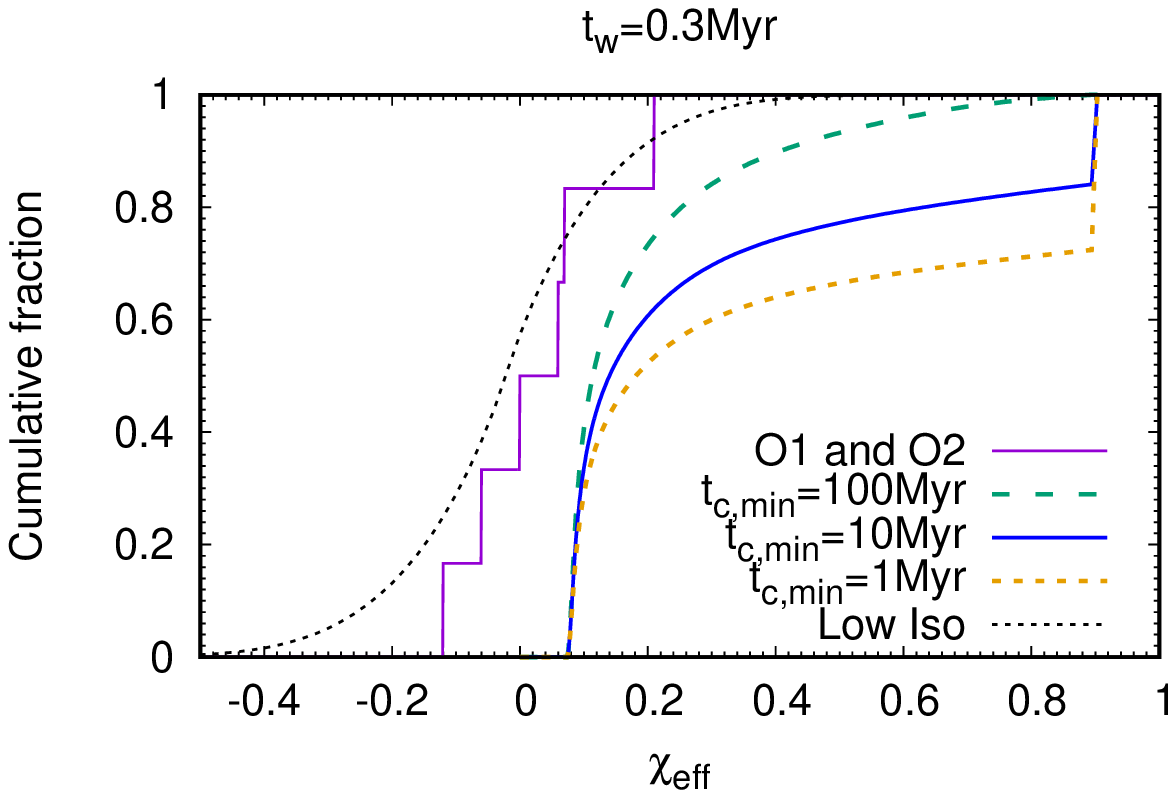}
         \includegraphics[width=6cm]{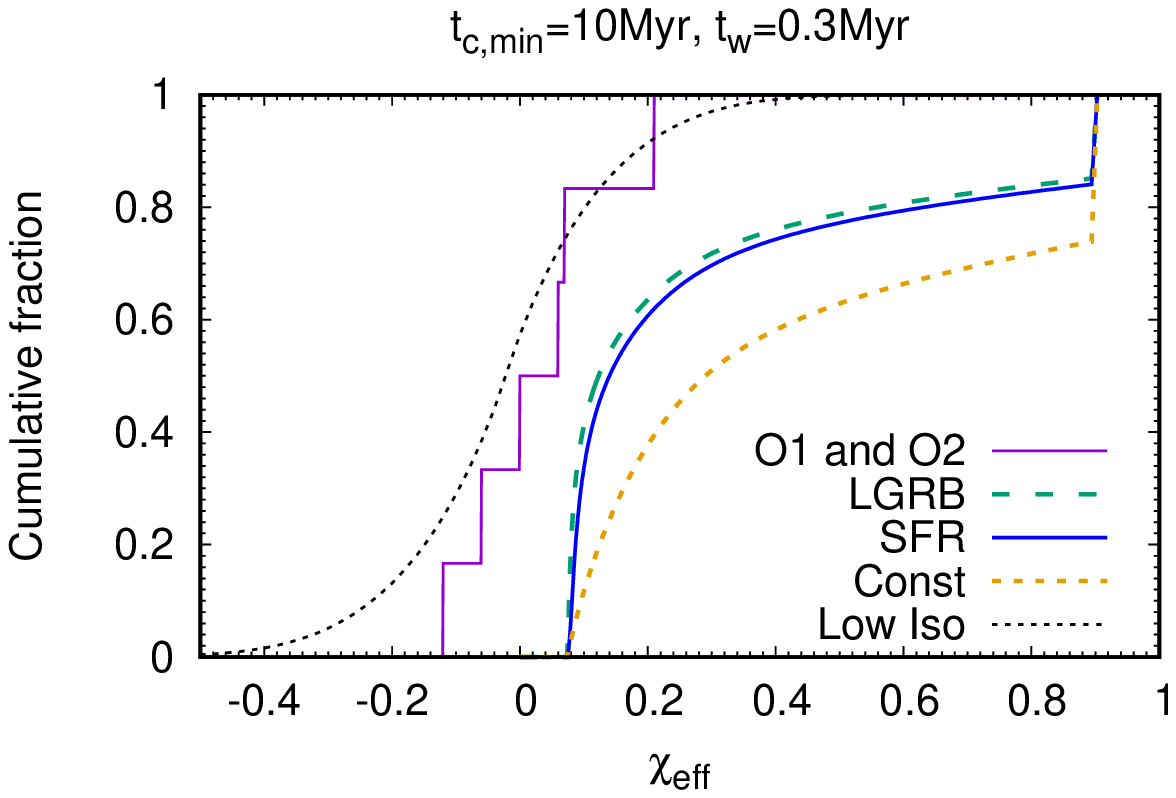}
         \includegraphics[width=6cm]{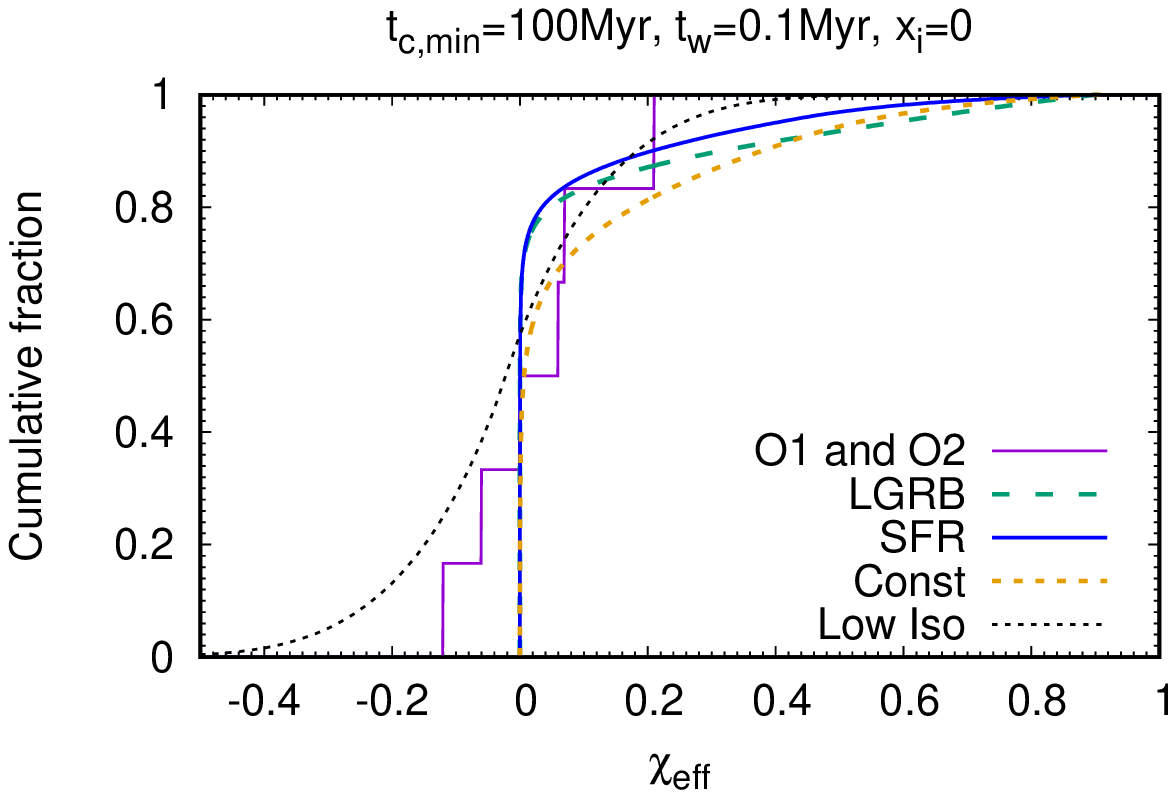}
         \includegraphics[width=6cm]{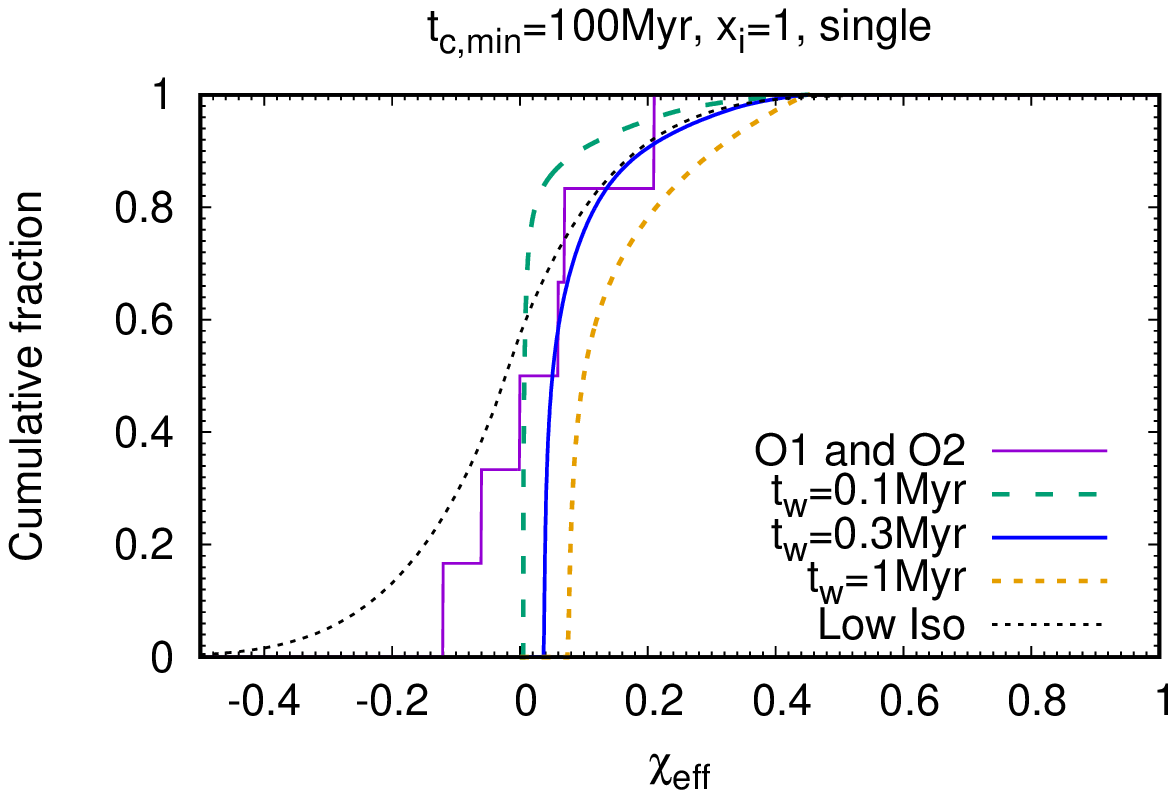}
           \caption{The cumulative $\chi_{\rm eff}$ distribution  for the O1 and O2 observing runs  for the WR binary scenario with different parameters. For the fiducial model (a blue solid line in the four upper panels), 
   the  BBH formation history  follows the cosmic SFR, the two stars are synchronized at the beginning of the WR phase,  the merger
   delay-time distribution is $\propto t^{-1}$ with a minimal time delay  of $t_{c,\,{\rm min}}=10$~Myr and the wind timescale is $t_{\rm wind}=0.3\,$Myr.  
   We set the mass ratio, $q=1$, and  $m_{\rm tot} = 60M_{\odot}$ for all models.  Also shown as a black dashed curve is the low-spin isotropic model in Ref. \citenum{farr2017Nature}. The two bottom panels show models that deviate significantly  from this fiducial choice (e.g. $t_{c,\,{\rm min}}=100$~Myr). In the left bottom panel the WR stars are not synchronized initially ($\chi_i=0$) . With a  strong wind and a long merger time delay distribution that follow the SFR or LGRB rates these systems produce very narrow low $\chi_{\rm eff}$ distributions. A constant formation rate gives here a better fit to the data. In the right bottom panel the stars are initially synchronized with  a very strong wind and the long delay time the distribution is consistent with the observations. 
%   Note also that the theoretical curves are  calculated for BBHs with a total mass of $60M_{\odot}$. 
}
    \label{fig:KS}
  \end{center}  
\end{figure*}

\begin{figure*}
  \begin{center}
        \includegraphics[width=6cm]{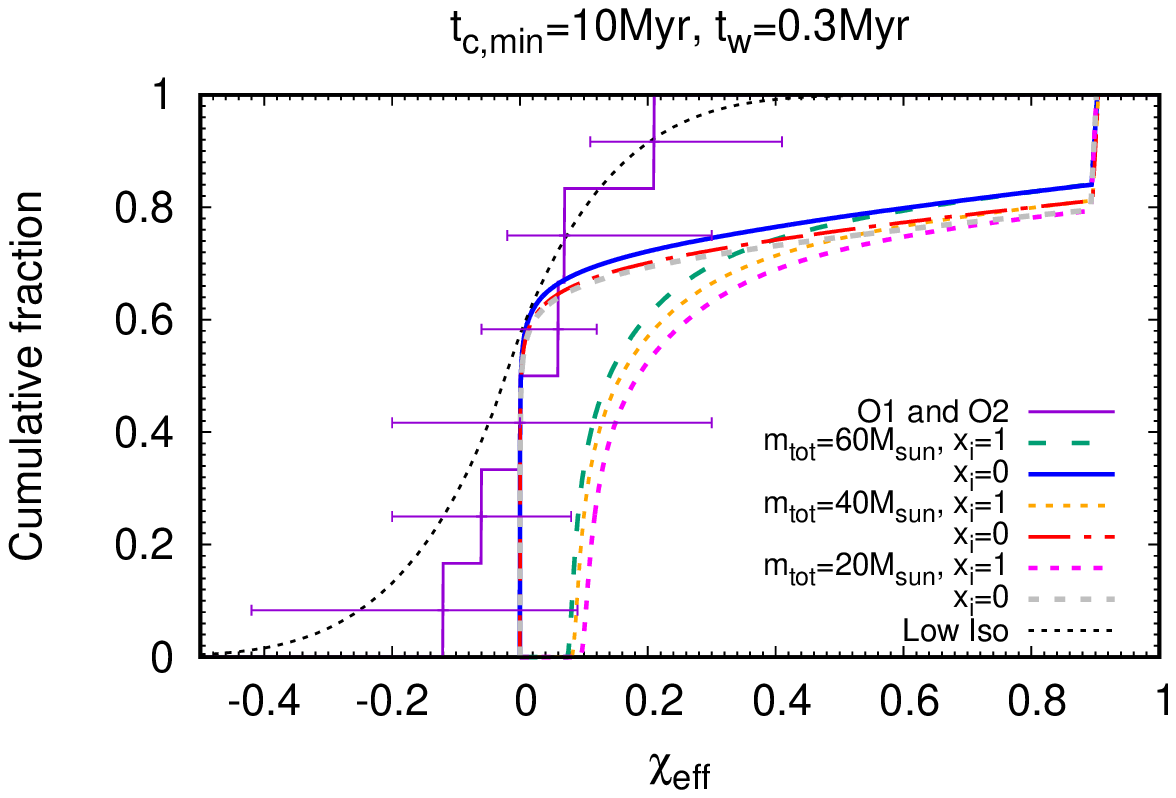}
         \includegraphics[width=6cm]{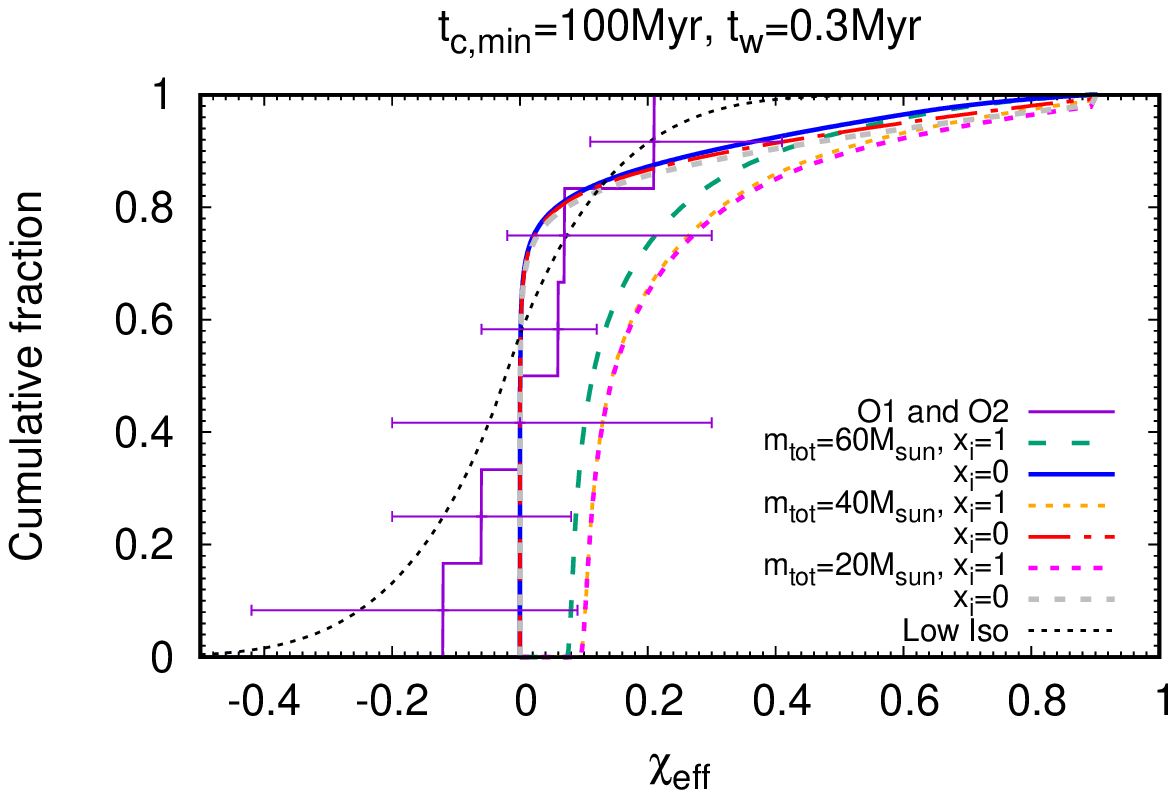}\\
                   \includegraphics[width=6cm]{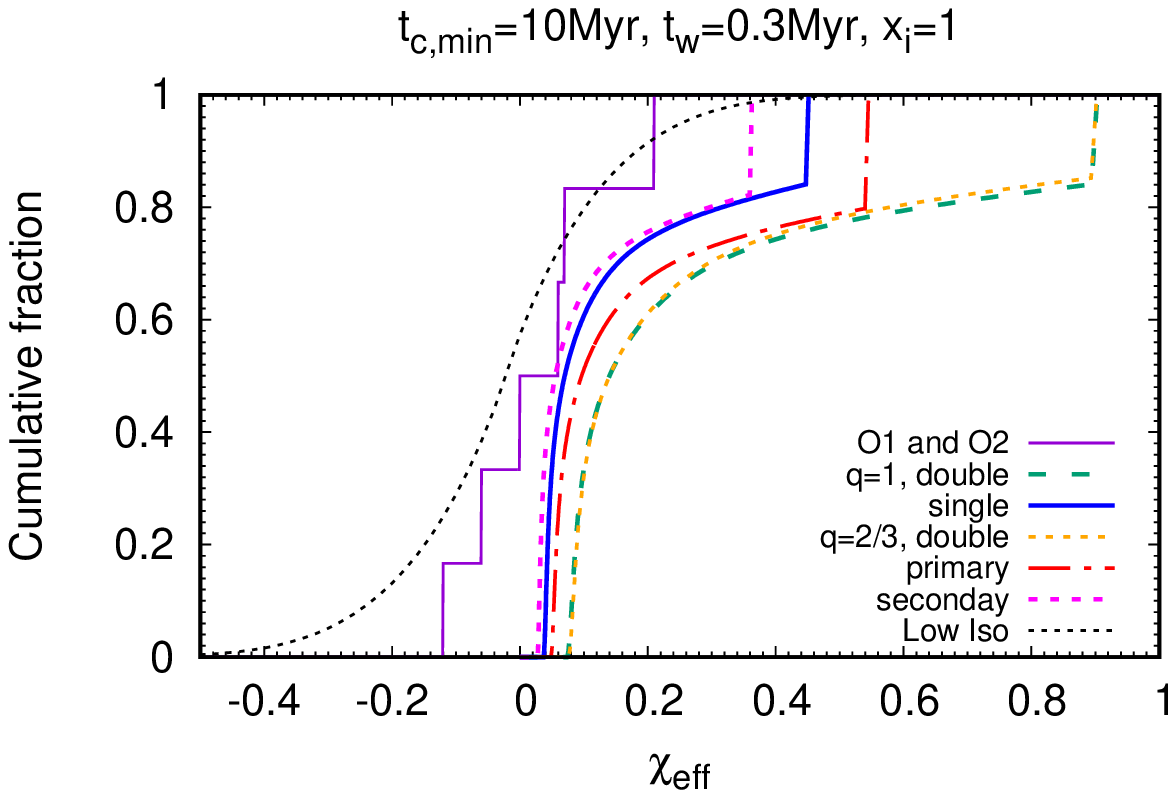}
         \includegraphics[width=6cm]{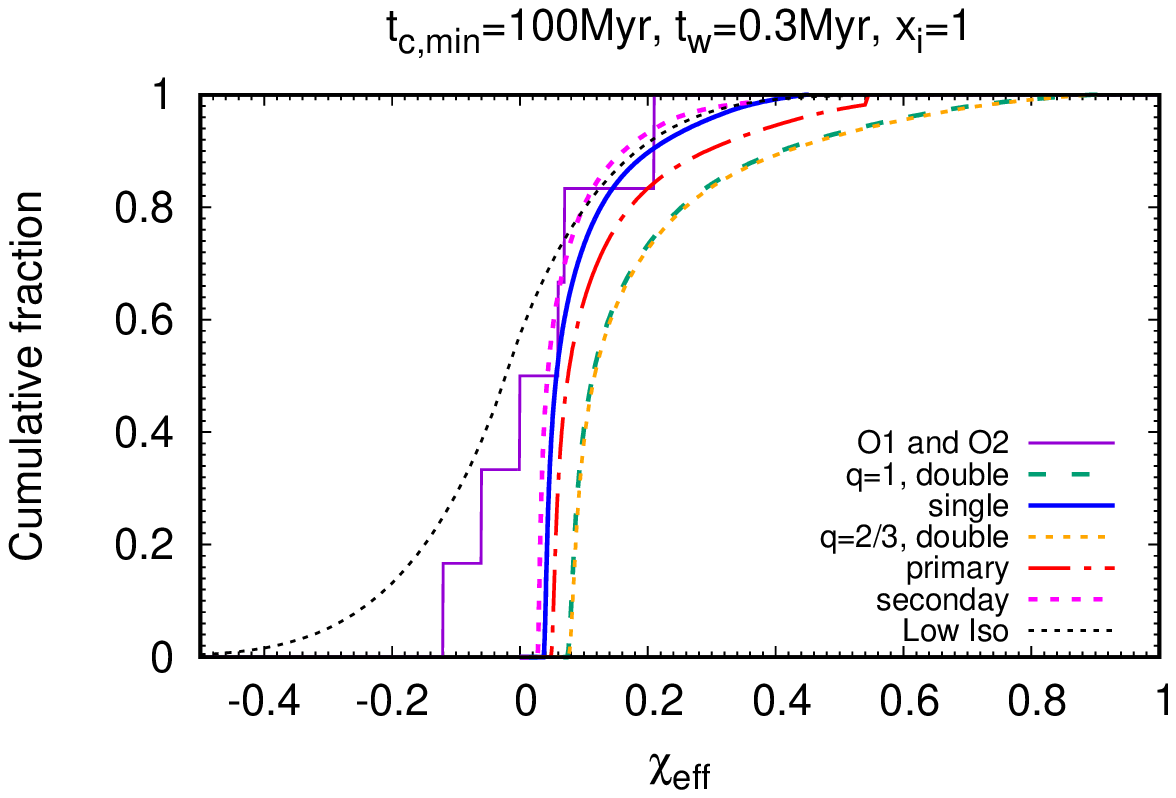}\\
                            \includegraphics[width=6cm]{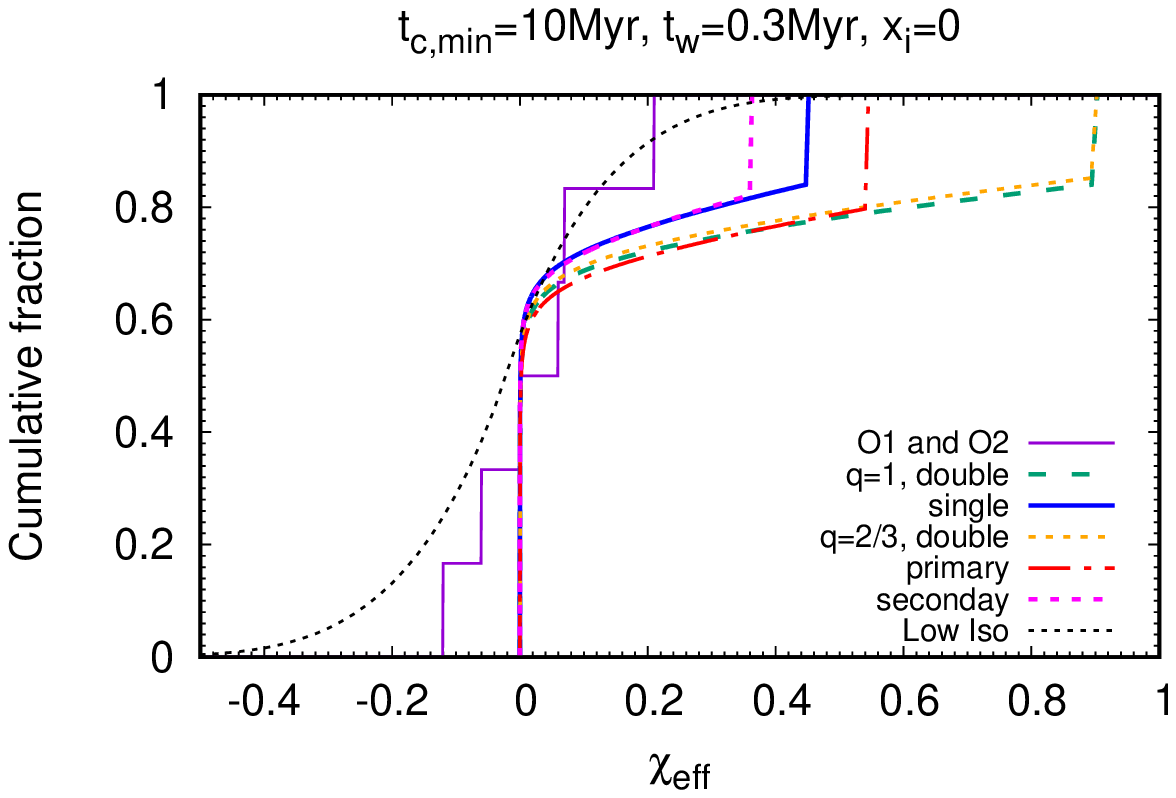}
         \includegraphics[width=6cm]{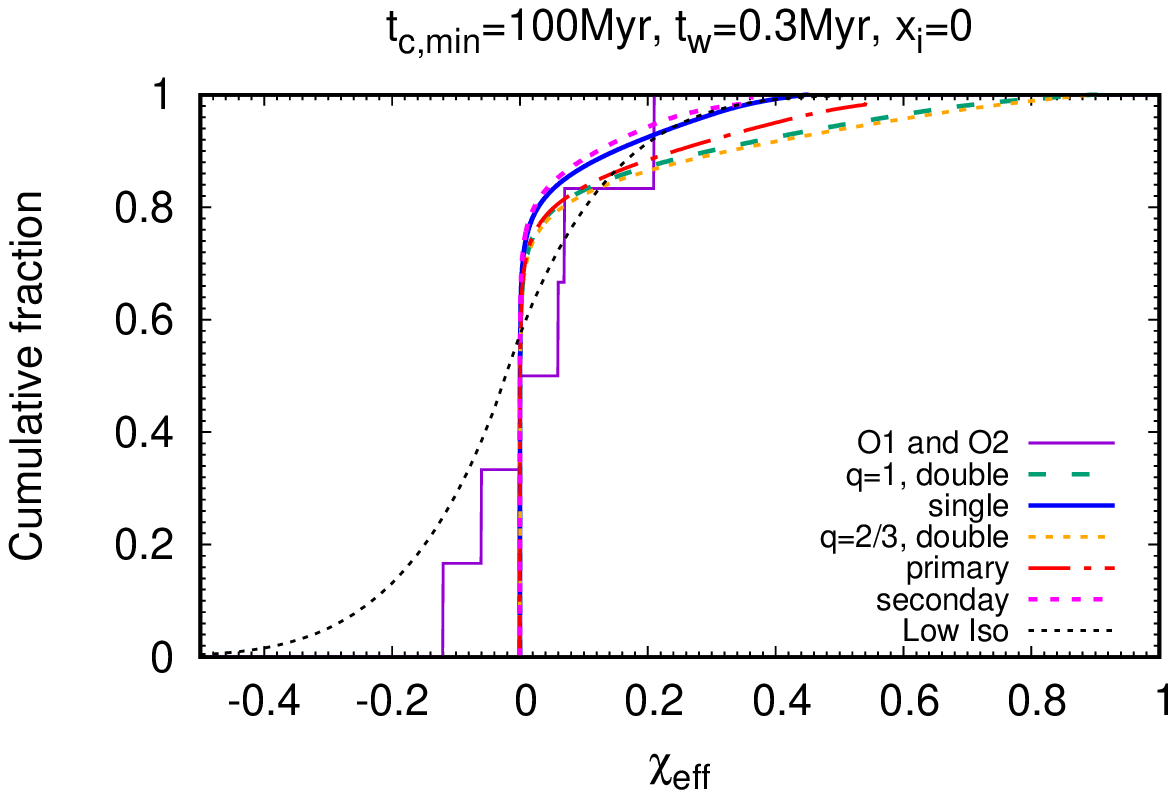}\\
           \caption{The same as Fig. \ref{fig:KS} but for different total masses and mass ratios. Here we use
           a BBH formation history proportional to the cosmic star formation history. In the middle and bottom panels, the curves labeled by ``primary" (``secondary") show  models
           in which one of the star i.e.  the primary (secondary)  is tidally synchronized. }
    \label{fig:KS2}
  \end{center}  
\end{figure*}

\section{A comparison with observations and implications} 
\label{sec:results}

%\cite{HotokezakaPiran2017ApJ} 
%\deltp{examined  the expected $\chi_{\rm eff}$ for BBH binaries of different types (see their Fig. 2). Either red or blue giant 
%progenitors are easily ruled out. Even regular main sequence stars would lead to progenitors' spin values that are much larger than unity. The only possible candidates are Pop III stars, that have in fact been predicted to produce massive BBHs  
%\citep{kinugawa2014MNRAS} and 
%WR stars, stripped massive stars that have lost their H and He envelopes.}
 
The observed spin values are low (consistent with zero) even when compared with those expected for WR stars \citep{HotokezakaPiran2017ApJ,zaldarriaga2017}. The ``tension" appeared already in the first detection  and it was intensified with the additional observations of low $\chi_{\rm eff}$ values and in particular with the observation of GW170104.  

To quantify  this issue we turn now to  estimate the expected  $\chi_{\rm eff}$ distribution of BBHs arising from WR binaries (see Refs. \citenum{zaldarriaga2017,HotokezakaPiran2017ApJ}) and compare it with the observations.
 %that are common at the final stage of evolution in the classical field binary scenario.  
 The distribution depends on the formation rate of these binaries. 
 We consider 
 BBH formation rates that follow the cosmic star formation rate (SFR) \cite{madau2014ARA&A} and a constant BBH formation rate. Since WR stars are also considered to be the progenitors of type Ic Supernovae that accompany LGRBs 
 we consider the possibility that the BBH formation rate follows the rate of Long GRBs (LGRBs).  In fact,  the rate of LGRBs is comparable to the rate of BBH mergers and an interesting possibility is that LGRBs accompany the formation of the massive BH that compose the  BBH that eventually merge.  
 %. To form massive BHs evolutionary scenarios require  low metallicity  progenitors  (otherwise the mass loss would be very significant). LGRBs are known to arise preferably in low metallicity hosts.
 %\footnote{From a theoretical point of view it has been suggested thatstrong winds that arise in higher metallicity progenitors will prevent fast rotation which is probably needed to produce a LGRB.}.  
Therefore we also consider  a BBH formation history that follows the long GRB rate \citep{wanderman2010MNRAS}. 
 The resulting spin distribution arising from of the SFR model is not very different from the one that follows LGRBs. On the other hand,
 the results for a constant BBH formation rate are quite different as a constant rate produces a significant fraction of  binaries formed at low redshifts. To merge sufficiently  rapidly those systems must have  small initial separations resulting in high spin mergers.

We expect that the rate of  mergers  follows the BBH formation  rate with a time delay $t_c$ whose probability  is distributed  as $\propto t_c^{-1}$. 
We consider a minimal time delay of 1 Myr  (corresponding to an initial separation of $3 \cdot 10^{11}$cm  for a $\approx 30 m_\odot$ BBH), 10 Myr ($5.4 \cdot 10^{11}$cm), or 100 Myr ($ 10^{12}$cm) between the formation of the BBH and its merger. These differences are important as the synchronization time depends strongly on the separation and hence on $t_c$. 
We also consider different wind timescales,  $t_{\rm wind}$: 0.1, 0.3 and 1 Myr. 
Note that the shortest time scale considered, $t_{\rm wind}=0.1$ Myr, roughly corresponds to a mass loss of  $10^{-4.5} m_\odot$/yr  which is at the level of the strongest winds  observed in WR stars~\citep{crowther2007ARA&A,vink2017}.
 With these assumptions we obtain several probability distributions for the observed $\chi_{\rm eff}$ values.
{In general, the field binary scenario predicts a bimodal $\chi_{\rm eff}$ distribution with low and high spin peaks\footnote{The low and high spin peaks are around $\chi_{\rm eff}\approx0$--$0.1$ and $\approx 1$, respectively.} (see Ref. \citenum{zaldarriaga2017,HotokezakaPiran2017ApJ} for simple models and Ref. \citenum{belczynski2017,postnov2017arXiv} for population synthesis studies).
 Here the high spin peak corresponds to tidally synchronized binaries.  }
 
 %Fig. \ref{fig:spin2} shows an example of such a comparison for a WR progenitor model. Once more the discrepancy between the observed values and the expectations is apparent. 

  Figure \ref{fig:KS}  depicts  the integrated observed distribution of $\chi_{\rm eff}$ compared with several WR models.  One can see the variety of the resulting $\chi_{\rm eff}$ distribution: some models  give $30$--$40\%$ of high $\chi_{\rm eff}$ mergers, while  $\chi_{\rm eff}$ is concentrated around zero for others. 
  The   models with the lowest  $\chi_{\rm eff}$ distributions are those in which the progenitors: 
 (i) Are not synchronized at the beginning  of the WR phase ($\chi_i=0)$; 
(ii)  Have a strong ($t_{\rm wind}=0.1$ Myr)  wind\footnote{Note however that such winds might not be consistent with very massive remnants.} ;
 (iii) Have a long ($t_{c,{\rm min}} = 100$ Myr) minimal time delay  - corresponding to a large initial separation.  
The question whether one or two of the  progenitors is influenced by the tidal interaction is secondary as it determines the largest $\chi_{\rm eff}$  values ($> 0.4$ or $>0.8$) and those have  not been observed so far. 
  
 Models with $\chi_i=1$, a  moderate wind ($t_{\rm wind} = 0.3$Myr) and a long  delay (100 Myr) in which the BBH formation rate follows the  SFR or LGRBs rates are consistent with the observations (apart from the nominal negative values, of course, but those could be due to the large measurement errors). 
  %Similarly $\chi_i=1$, a strong wind  ($t_{wind} = 0.1$Myr ) and a long (100 Myr) delay with a constant BBH formation rate are also consistent. 
  The top four panels compare different models to a fiducial model in which both progenitors are spinning rapidly at the beginning of the WR phase $\chi_i=1$,  $t_{\rm wind}=0.3$ Myr, $t_{c,{\rm min}}=10 $Myr, and the BBH formation rate following the cosmic SFR. 
The lower two panels depict  more extreme models. Here we find that if all the above conditions are satisfied then the resulting $\chi_{\rm eff}$ distribution (for SFR or LGRB rate) is too narrowly concentrated around zero. A better fit to the data is obtained under these conditions if the BBH formation rate is a constant (see bottom left panel of Fig. \ref{fig:KS}). Even if the progenitor stars are synchronized at the beginning of the WR phase ($\chi_{i}=1$) a  strong enough wind ($t_{\rm wind}=0.1$ Myr) can lead to sufficient angular momentum loss  so that the final $\chi_{\rm eff}$ distributions would be concentrated close to zero (see bottom right panel of Fig. \ref{fig:KS}). 

{Figure \ref{fig:KS2} shows the dependence of the $\chi_{\rm eff}$ distribution on the total mass  and the mass ratio.
Note that the dependence on these parameters is rather weak except for the cases in which  one of the component stars is 
synchronized (singly synchronized). This is because of the dependence of Eq. (\ref{eq:synWR}) on the mass ratio.
These differences do not change qualitatively our conclusions. }

{In order to take the relatively large measurement errors of $\chi_{\rm eff}$ into account when comparing 
different models with the observed data, we evaluate the odds ratios between the marginalized likelihoods of different field evolution  models 
and the low-isotropic spin model of   Ref. \citenum{farr2017Nature}, the most favorable one among the synthetic models  considered.} 
We calculate the marginalized likelihood of each model for the six events, $p_i(d|M)$ and then combine them as $p(d|M)=\prod_{i}p_i(d|M)$. Tables 2 and 3 list the odds ratios of these different models of equal mass binaries with $m_{\rm tot}=60M_{\odot}$
to the low-isotropic spin model, $p(d|M)/p(d|{\rm Low\,Iso})$. 
The field binary evolution models with a long delay time ($100$\,Myr) and a strong wind ($t_{\rm wind}=0.1$\,Myr) 
have an odds ratio of about unity.  
% so that low-isotropic spin model is more consistent with the observed $\chi_{\rm eff}$ distribution than our aligned WR binary models. }
Furthermore, many of the models satisfying the conditions mentioned above have $p(d|M)/p(d|{\rm Low\,Iso})\gtrsim 0.1$. In spite of the two observed cases with negative mean $\chi_{\rm eff}$ values,  these models cannot be ruled out with the current $\chi_{\rm eff}$ distribution of six observed events. 
%%%%Therefore we conclude that the field binary models with a long delay and a strong wind and the low-spin dynamical capture model are both consistent with the observed distribution of $\chi_{\rm eff}$. We will be able to obtain a much stronger conclusion  when we have a few more events.

\begin{table*}
\begin{center}
{Table 2: Odds ratio of the models to the low-isotropic spin model for  initially synchronized WR binaries and double (single) synchronization.}
\label{tab:odds1}
\vskip0.5cm
%\scalebox{1.}
{\begin{tabular}{lccc}
\hline \hline
Model ($t_{c,\,{\rm min}}$) & $t_{\rm wind}=0.1$~Myr           &   $t_{\rm wind}=0.3$~Myr               & $t_{\rm wind}=1$~Myr                 \\  \hline
SFR (1Myr) & $0.14~(0.24)$ & $0.07~(0.30)$ & $<0.001~(0.08)$ \\
LGRB (1Myr) & $0.16~(0.26)$ & $0.1~(0.36)$ & $0.001~(0.12)$  \\
Const (1Myr) & $0.04~(0.09)$ & $0.003~(0.06)$ & $<0.001~(0.004)$                         \\ \hline
SFR (10Myr) & $0.34~(0.51)$ & $0.17~(0.70)$ & $0.001~(0.21)$  \\
LGRB (10Myr) & $0.36~(0.52)$ & $0.24~(0.78)$ & $0.003~(0.29)$ \\
Const (10Myr) & $0.14~(0.30)$ & $0.01~(0.20)$ & $<0.001~(0.02)$ \\ \hline
SFR (100Myr) & $1.05~(1.23)$ & $0.52~(2.00)$ & $0.004~(0.70)$  \\
LGRB (100Myr) & $1.02~(1.17)$ & $0.68~(2.06)$ & $0.01~(0.89)$  \\
Const (100Myr) &$ 0.97~(1.56)$ & $0.08~(1.28)$ & $<0.001~(0.14)$  \\ 
\hline \hline 
\end{tabular}}
\end{center}
%\caption{Odds ratio of the models to the low-isotropic spin model for  initially synchronized WR binaries and double (single) synchronization.}
\end{table*}

\begin{table*}
\begin{center}
{Table 3: Odds ratio of the models to the low-isotropic spin model  for  initially non-rotating WR binaries.}
\vskip 0.5cm
\label{tab:odds2}
%\scalebox{1.}
{\begin{tabular}{lccc}
\hline \hline
Model ($t_{c,\,{\rm min}}$) & $t_{\rm wind}=0.1$~Myr           &   $t_{\rm wind}=0.3$~Myr               & $t_{\rm wind}=1$~Myr                 \\  \hline
SFR (1Myr) & $0.11~(0.22)$ & $0.09~(0.19)$ & $0.08~(0.22)$ \\
LGRB (1Myr) & $0.12~(0.23)$ & $0.10~(0.21)$ & $0.09~(0.23)$  \\
Const (1Myr) & $0.03~(0.08)$ & $0.02~(0.06)$ & $0.01~(0.08)$                         \\ \hline
SFR (10Myr) & $0.27~(0.47)$ & $0.22~(0.42)$ & $0.20~(0.38)$  \\
LGRB (10Myr) & $0.28~(0.48)$ & $0.24~(0.44)$ & $0.22~(0.40)$ \\
Const (10Myr) & $0.11~(0.28)$ & $0.08~(0.22)$ & $0.07~(0.18)$ \\ \hline
SFR (100Myr) & $0.80~(1.10)$ & $0.69~(1.13)$ & $0.65~(1.07)$  \\
LGRB (100Myr) & $0.79~(1.06)$ & $0.68~(1.09)$ & $0.64~(1.04)$  \\
Const (100Myr) & $0.79~(1.43)$ & $0.57~(1.31)$ & $0.50~(1.19)$  \\ 
\hline \hline 
\end{tabular}}
\end{center}
\caption{Same as Table 2 but for  initially non-rotating WR binaries.}
\end{table*}

\section{Conclusions}
\label{sec:summary} 

Before discussing the implications of these findings
 we turn, once more, to possible caveats. We have already argued that observations of Galactic X-ray binaries including massive BHs provide a good evidence for our model for the formation of massive BHs (no kick and no mass loss). 
The main open issues are all related to the late phases of the stellar binary evolution of very massive stars:
% \begin{romanlist}[(iii)]
 %\item 
 (i) What are  the spins of the BHs at the end of the common envelope phase? 
%\item 
(ii) What is the separation at the end of the common envelope phase? 
%\item 
(iii) How strong are the winds? 
% \end{romanlist}
 The answers to these questions depend on better understanding these late phases. 
 
Turning now to the results, 
the negative observed values are clearly inconsistent with the  evolutionary model (unless there are significant kicks at the formation of the BHs\cite{Bekenste73}).
However, while the observed low $\chi_{\rm eff}$  values are at some ``tension" with the expectations of the standard evolutionary scenario (see also Ref. \citenum{kushnir2016MNRAS,HotokezakaPiran2017ApJ,zaldarriaga2017}),
the large error bars of the $\chi_{\rm eff}$ measurements don't allow us to rule out some of the field binary scenario models. 
%It is  also interesting that none of the six events observed implies  (at $> 2\sigma$) a negative $\chi_{\rm eff}$. 
These large errors combine with  the small number statistics makes it impossible to make now clear conclusions.  

%More important  are the uncertainties in the outcome of the common envelope phase (whether the stars are synchronized or not at the end of this phase), in the strength of the wind at the late stages of the evolution of these massive stars, in the minimal time delay for mergers (corresponding to the minimal separation and hence to the importance of the tidal locking process) and finally in the tidal synchronization process itself. 
%For example, it is clear that strong enough winds would reduce the final spin of the stars. However, one may wonder if such strong winds are consistent with the very massive BHs observed. 
%Note that $t_{\rm wind}=0.1$ Myr roughly corresponds to a mass loss of  $10^{-4.5} m_\odot$/yr  which is at the level of the strong winds of the observed WR stars~\citep{crowther2007ARA&A,vink2017}. 

%If two or more of these uncertain processes (e.g. longer time delays and stronger winds) tends towards a more extreme values.

Both the SFR and  LGRB rates are favorable as proxies for the BBH formation rate. In both cases most of the formation takes place at early times, allowing for large initial separations. The resulting $\chi_{\rm eff}$ distributions arising from these two scenarios are practically indistinguishable. A constant BBH formation rate implies more recent formation events and hence shorter merger times leading to  larger   $\chi_{\rm eff}$ values. Still with 
extreme parameters even this distribution can be made consistent with the current data. 

%Furthermore the synchronization and tidal locking process are not sufficiently clear and it might be that some stars simply are not locked. Still, it is interesting to note on one hand some conclusions and on the other to make prediction for the spin values, that will be observed in the future within the evolutionary scenario. 
% \deltp{the only ``reasonable" progenitors within this scenario are WR  or Pop III stars. The former are 
% more compact, the latter live at very high redshifts and hence they naturally have longer merger times.} 
WR stars formed at high redshifts, $z\gsim 2$, are the best candidates for being progenitors with low $\chi_{\rm eff}$. They have a long merger time,  allowing them to begin with a relatively large separation that implies much weaker synchronization.   
 However, this is not enough and strong or moderate winds (for  progenitors that are non-rotating at the end of the common envelope phase)  are essential  for consistency with the current distribution. A longer  minimal time delay (corresponding to larger separations at the birth of  BBHs) helps, but is insufficient to lead to consistency. 
  
% \deltp{With the same reasoning 
% according to this model, BBHs that merge earlier (at higher redshifts) and  have had less time to merge. This implies that they have originated from systems with smaller semi-major axes and hence  their projected spins must be larger. Thus a more sensitive detector that will be able to detect BBH mergers from higher redshifts must see higher spin systems. }

To conclude  we note that a comparison of the currently observed O1 and O2 $\chi_{\rm eff}$ values with the models show some tension, however it does not rule out evolutionary models based on WR stars.  In fact some  models are almost as consistent as the best fitted low-isotropic spin model of Ref. \citenum{farr2017Nature}. While many models predict a significant fraction ($>25\%$) of large ($>0.4$ for singly synchronized and  $>0.8$ for double synchronization) $\chi_{\rm eff}$ events, some  produce distributions that are concentrated around positive very low $\chi_{\rm eff} $ values.  The question which models are consistent depends on largely unexplored late stage evolution of very massive stars. 
Given these results it seems that while a significant fraction of high $\chi_{\rm eff}$ mergers will strongly support the field evolutionary scenario, lack of those will be hard to interpret. It may indicate another scenario or, for example,   strong winds that remove the spin angular momentum. On the other hand a significant fraction of negative $\chi_{\rm eff} $ merger will be difficult to reconcile with this scenario, unless the BHs' angular momentum is dominated by very strong natal kicks. 
  
%Before concluding we turn now to discuss a few caveats that may influence our conclusions and release the apparent ``tension" and the binary evolution model. Some of these issues have been discussed earlier. Still we repeat them here fro clarity. 
%\begin{Itemize} 
%\item The first caveat involves the assumption that the BBH has a  circular orbit. A highly  elliptical orbit can reduce significantly the merger time ???
%\end{itemize}

\section*{Acknowledgments}

KH is supported by the Flatiron Fellowship at the Simons Foundation and
Lyman Spitzer Jr. Fellowship at Princeton University.
TP is supported by an advanced ERC grant ``TReX" and by the ISF-CHE I-Core center of excellence for research in Astrophysics.

\end{document}